\documentstyle[epsf]{article}

\begin{document}
\newcommand{\beq}{ \begin{equation} }
\newcommand{\eeq}{ \end{equation} }
\newcommand{\bea}{ \begin{eqnarray} }
\newcommand{\eea}{ \end{eqnarray} }
\newcommand{\be}{ \beta }
\newcommand{\f}{ \frac }

\thispagestyle{empty}
\parskip=12pt
\raggedbottom

\def\mytoday#1{{ } \ifcase\month \or
 January\or February\or March\or April\or May\or June\or
 July\or August\or September\or October\or November\or December\fi
 \space \number\year}
\noindent
\hspace*{9cm} IFUP-TH 23/95\\
\vspace*{1cm}
\begin{center}
{\LARGE Scaling and topology  in the 2--d O(3) $\sigma$--model \\
  on the lattice with the fixed point action}\footnote{Work supported in 
part by Fondazione ``A. Della Riccia'' (Italy).} \\

\vspace{.5cm} 
(Revised version) \\
\vspace{1cm}

Federico Farchioni, Massimo D'Elia \\
Dipartimento di Fisica dell'Universit\`a and I.N.F.N.\\
Piazza Torricelli 2, I-56126 Pisa, Italy. 

\vspace{.5cm}

Alessandro Papa\footnote{On leave from the Department of Physics, University 
of Pisa and I.N.F.N., Pisa.} \\

Institute for Theoretical Physics \\
University of Bern \\
Sidlerstrasse 5, CH-3012 Bern, Switzerland

\vspace{0.5cm}

\mytoday \\ \vspace*{0.5cm}

\nopagebreak[4]

\begin{abstract}
We  study scaling properties and  topological aspects 
of the 2--d O(3) non--linear $\sigma$--model on the lattice with the 
parametrized fixed point action recently proposed  by P.~Hasenfratz  
and F.~Niedermayer. 
The behavior of the mass gap confirms the good properties of scaling of the
fixed point action.
Concerning the topology, lattice classical solutions are proved 
to be very stable under local minimization of the action; this outcome 
ensures  the reliability of the cooling method for the computation 
of the topological susceptibility, which indeed reproduces the 
results of the field theoretical approach.
Disagreement is instead observed with a different
approach in which the fixed point topological charge operator is used: 
we argue that the discrepancy is related  to the 
ultraviolet dominated nature of the model.
\end{abstract}

\end{center}
\eject

\pagestyle{plain}
\pagenumbering{arabic}

\section{\bf Introduction}
\label{sec:intro}

Simulating a theory on the lattice is the most (and often the only) reliable
mean to get an insight into its  non--perturbative aspects.
Once the theory has been put on the lattice, the non--trivial question is
how to correctly reach the continuum limit. The standard procedure of 
adjusting the  bare parameters of the lattice theory closer and closer 
to the critical values (where the lattice theory is expected to reproduce the
continuum) is plagued by critical slowing--down and finite--size effects: in 
practice one is forced to simulate the theory at finite cut--off, where the 
systematic effects of discretization, heavily coming into play, can be 
removed only by an -- often ambiguous -- extrapolation. A way to reduce 
lattice artifacts is to use perturbatively improved lattice 
actions~\cite{Sym80}, whose performance in numerical simulation is not, 
however, under theoretical control.
 
A radical solution to the problem~\cite{HN94} is to use perfect actions,
i.e. lattice actions whose spectrum is completely free of lattice artifacts. 
Wilson's renormalization group theory~\cite{Wil74} ensures that the fixed 
point (FP) of a renormalization group (RG) transformation and all 
lattice actions of the renormalized trajectory (RT) -- which is the 
asymptotic flux line under repeated RG transformations -- are perfect 
(quantum) actions in the above mentioned sense. The FP action, in 
particular, represents the classical perfect action, having 
the same classical properties of the continuum theory~\cite{HN94}.

A method for the determination of the FP of asymptotically free 
theories has been proposed in a recent paper~\cite{HN94}. The procedure
has been applied to the 2--d O(3) non--linear $\sigma$--model on the 
lattice, and a parametrization of the FP action $A_{FP}$ suitable for
numerical simulations has been found. Strong numerical evidences~\cite{HN94}
indicate that the action $A_{FP}$ is a good 
approximation of the perfect action at small correlation lengths also.
This agrees with the general remark by Wilson~\cite{Wil80} that the FP action
is (quantum) perfect at 1--loop order in perturbation theory. A formal
argument for this statement is presented in~\cite{DHHN95} and a further 
support comes from a specific 1--loop calculation in the O(3) $\sigma$--model
\cite{FHNP95}.

In the first part of this paper, we test the scaling properties of the FP 
action of Ref.~\cite{HN94} up to very small correlation 
lengths: we study  the physical scaling of the mass gap  
-- which is the only relevant spectral property in the O(3) $\sigma$--model 
-- by comparing different lattice definitions; we also check the rotation 
invariance.

The second part of this work is devoted to the study of the topological 
properties of the model with the FP action: in particular, we address 
ourselves to the problem of extraction from the lattice of the topological 
susceptibility of the O(3) 
$\sigma$--model~\cite{BL81}\cite{Lue82}\cite{Ber81}\cite{MPV82}\cite{BP85}\cite{DFPV92}\cite{BBHN94}\cite{AB95}. We present different 
approaches to the subject -- the field theoretical~\cite{FTM}, the 
geometrical~\cite{BL81} and the cooling~\cite{Ber81}\cite{Tep86} methods -- 
and compare their outcomes. 
In this respect, we argue that the cooling method can be safely applied
with the FP action, since it possesses scale invariant classical 
solutions~\cite{HN94} and no loss of topological signal is expected during 
the procedure of minimization. The scaling of topological susceptibility is a
rather involved matter since the semiclassical 
approximation~\cite{WKB}\cite{RR83} and some numerical 
evidences~\cite{MS94}\cite{FP94} indicate that the topology of the model is 
UV dominated.

\section{\bf The model}
\label{sec:model}

The  2--d O(N) $\sigma$--models are O(N)--symmetric renormalizable 
quantum spin field theories, asymptotically free for N$\geq3$.

They are described by the Lagrangian:
\beq
{\cal L} = \frac{\beta}{2} \: \partial _{\mu} \phi(x) \cdot \partial _{\mu} 
\phi(x) \;\;\;\; ,
\label{eq:def}
\eeq
where $\phi (x)$ is a N--component real field satisfying the constraint
$\phi \cdot \phi = 1$ and $\be$ is the inverse of the coupling constant $g$.

The infrared charge singularity of these models is responsible for 
the disintegration of the Goldstone vacuum~\cite{BZ-JLeG76};  for N$\geq 3$ 
the true vacuum is O(N) symmetric and non--degenerate.
In the limit N$\rightarrow\infty$ the model contains an isovector
N--plet of free massive particles only~\cite{BZ-J76}\cite{BLS76}. 
The interaction of these particles is of order 1/N and so for N 
sufficiently large no bound states are
present. Strong theoretical evidences bear the conjecture 
that the situation is the same at all N $\geq3$~\cite{Z-Z78}.
Consequently, the spectrum of the O(3) $\sigma$--model is expected to 
consist of a  single triplet of particles.

Among the  O(N)  $\sigma$--models, the O(3) model in particular plays an 
important role in quantum field  theory because it resembles, besides 
asymptotic freedom and spontaneous mass generation, another aspect of the 
4--d non--Abelian gauge theories, i.e. the non--trivial topology.
 
The topological charge $Q$ of a 3--component spin field $\phi (x)$
is the number of times $\phi (x)$ winds the sphere $S^{2}$. 
It can be expressed as the integral over the space--time of a local 
operator $Q(x)$:
\beq 
Q(x) = \frac{1}{8\pi} \epsilon_{\mu \nu }
\epsilon_{ijk} \phi_{i} (x) \partial _{\mu} \phi_{j}(x) \partial _{\nu} 
\phi_{k}(x)  \;\;\;\;\;  ;
\eeq
$Q(x)$ is the divergence of a topological current $K_{\mu}$ 
\cite{tH76}\cite{DLD78},
\beq
Q(x)=\partial _{\mu }K_{\mu }(x) \;\;\; .
\eeq

All classical solutions with non--trivial topology -- the $k$--instantons -- 
have  been explicitly found~\cite{BP75}. At a quantum level, the only 
available prediction comes from the semiclassical approximation; the 1--loop 
result\footnote{Using the perturbative RG one can prove that this result is
in fact asymptotically exact as $\rho \rightarrow 0$\cite{Lue82}.} 
for the size 
distribution of instantons in the physical vacuum is $(1/V) dN/d\rho\propto 
1/\rho$. This expression gives a logarithmic divergence in the ultraviolet 
limit $\rho \rightarrow 0$.
Moreover, numerical approaches for the evaluation of the instanton size 
distribution~\cite{MS94}\cite{FP94} indicate a growing behavior when the 
size decreases (the lowest size investigated is $\rho\sim0.2\,\xi_G$).

The topological susceptibility is defined as the correlation at zero 
momentum of two topological charge density operators $Q(x)$~:
\beq
\chi \: = \: \int \: d^{2}x \;\langle 0|\:T\, [\:Q(x)Q(0)\:]\: |0 \rangle 
\:\:\: .
\label{eq:chicont}
\eeq

The prescription defining the product of 
operators in Eq.~(\ref{eq:chicont}) is~\cite{Cre79}
\beq
\langle 0|\: T [\:Q(x)Q(0)\:]\: |0\rangle\: \equiv \partial _{\mu }
\langle 0|\: T[\:K_{\mu }(x)Q(0)\:]\:  |0\rangle \: \: .
\label{eq:crew}
\eeq
This prescription eliminates the contribution of possible contact terms 
(i.e. terms proportional to the $\delta $ function or its derivatives) when
$x\rightarrow 0$.

\section{\bf Scaling: mass gap measurements}
\label{sec:mass--gap}

The general parametrization of the FP action of the O(3) $\sigma $--model is
\bea
A_{FP}(\phi)= \be \left\{\right.&\!\!\!\!-\!\!\!\!&\f{1}{2} \sum_{x,r}\rho(r)(1-\phi(x)\cdot\phi(x+r))
\\
&\!\!\!\!+\!\!\!\!&\sum_{x_1,x_2,x_3,x_4}c(x_1,x_2,x_3,x_4)
(1-\phi(x_1)\cdot\phi(x_2))(1-\phi(x_3)\cdot\phi(x_4)) + \;\;\; ... \;\;
\left.\right\} ,
\nonumber
\eea
where $\rho$ represents the perfect discretization of the 
Laplacian and all the (infinite) multi--spin couplings with more than 4
fields have been only implicitly indicated. In this paper we test the
properties of the 24--couplings parametrization reported in Table~4 of 
Ref.~\cite{HN94}, which contains only one--plaquette terms with no more than
8 fields.

This parametrized FP action  should reproduce the continuum
even at moderate values of the correlation length: cut-off effects on the 
spectral properties of the lattice theory should be absent, or at least 
strongly suppressed. This is confirmed by numerical evidences~\cite{HN94}. We
want to check this point further.

We have performed our test by comparing alternative lattice definitions of 
mass gap.
The standard definition is related with the large--distance
behavior of the wall--wall correlation function\footnote{A small cut-off 
effect survives in the following definitions of the lattice mass gap since 
the ordinary lattice field $\phi$ instead of the FP field operator 
$\phi^{FP}$ is used~\cite{HN94}\cite{DHHN95}.} 
\beq
G_{w}(y-x) \; = \; \f{1}{L}\sum_{x_1,y_1}G(x_1,x;y_1,y) \; \; \; ,
\eeq
where
\beq
G(x) \; = \; \langle \phi(x)\cdot \phi(0) \rangle \;\;\; .
\eeq
The expected large--distance behavior, including periodic boundary condition
effects, is:
\beq
G_{w}(x)\simeq\f{A_{w}}{2}\left[{\rm exp}\left(-\f{x}{\xi_{w}}\right)+  
{\rm exp}\left(-\f{L-x}{\xi_{w}}\right)\right]  \; \; \;
\mbox{for}\;\;\; \f{L}{2}>x \gg \xi_{w} \; \; \; .  
\eeq

It is also possible to define the diagonal wall--wall correlation function
\beq
G_{d}\left(\f{y-x}{\sqrt{2}}\right) \; = \; \f{\sqrt{2}}{L}\sum_{x_1,y_1}
G(x_1,x-x_1;y_1,y-y_1) \; \; \; ,
\eeq
whose large--distance behavior is 
\beq
G_{d}(x)\simeq\f{A_{d}}{2}\left[{\rm exp}\left(-\f{x}{\xi_{d}}\right)+  
{\rm exp}\left(-\f{L/\sqrt{2}-x}{\xi_{d}}\right)\right]  \; \; \;
\mbox{for}\;\;\; \f{L}{2\sqrt{2}}>x \gg \xi_{d} \; \; \; .  
\eeq
One of the scaling tests consists in verifying that 
the ratios $\xi_d/\xi_w$ and $A_d/A_w$ keep constant when $\be$ varies.
In addition, the comparison between $\xi_d$$(A_d)$ and $\xi_w$$(A_w)$
allows to test directly the rotation invariance: scaling violations are 
revealed  by deviations from 1 of the ratios $\xi_d/\xi_w$ and $A_d/A_w$.

An alternative definition of the correlation length comes from considering
the second moment of the correlation function
\beq
\xi_G^2 \; = \; \f{\int d^2x \: \f{1}{4} x^2 \, G(x)}{\int d^2x \: G(x)} 
\;\;\; .
\eeq
We use the following lattice definition of $\xi_G$:
\beq
\xi_G^{2}\; =\; \f{1}{4\sin^2(\pi/L)}\left[\f{\tilde{G}(0,0)}
{\tilde{G}(0,1)}-1\right]\; \; \; ,
\eeq
where $\tilde{G}(k)$ is the Fourier transform of $G(x)$, given by
\beq
\tilde{G}(k) \: = \: \frac{1}{L^2}\sum_{x,y}\langle \phi (x)
\cdot\phi (y) \rangle \exp\left [ i\frac{2\pi}{L}(x-y)\cdot k \right ]
\; \; .
\eeq
The zero component of $\tilde{G}(k)$ is by definition the magnetic 
susceptibility $\chi_m$.
In the scaling region the ratio $\xi_G/\xi_w$ must be a constant, 
scale--independent number. The ratio should be equal to 1 within 1\% 
\cite{CR91}.

Let's introduce the quantity $A_G = \chi_m\xi_G^{-2}\xi_w$. The 
adimensional ratio $A_w/A_G$ is scheme--independent in the scaling region. 
We expect $A_w/A_G \approx 1$, because the two--point functions should be 
almost saturated by the lowest energy state. 

We performed Monte Carlo (MC) simulations for several values of $\be$ 
corresponding to correlation lengths $\xi$ varying from $\sim 2$ to $\sim 34$
lattice units. We adopted a 4--hits Metropolis algorithm which is the only 
local algorithm available for actions with multi--spin interactions. The 
local  nature of the algorithm will be essential in the application of the 
heating techniques to be discussed later. The acceptance of the 
pseudo--random changes of the Monte Carlo has been adjusted at 50\%.

In Table~I  we report values of correlation length and correlation function 
coefficient obtained, for the three definitions, at various $\be$'s.
The fits to $G_w$ and $G_d$ have been performed choosing $x_{min} \simeq 
\xi$; we have checked that larger values of $x_{min}$ give results consistent
within errors. Table~II shows the ratios of these different definitions, 
analyzed using jackknife techniques. We can observe nice scaling and rotation
invariance: in the region of $\be$ between 0.8 and 1.0 ($\xi$ between $\sim 
5$ and $\sim 12$) the uncertainty is lower than 1\%. Moreover, with the same 
precision the ratios $\xi_G/\xi_w$ and $A_G/A_w$ are consistent with one, so 
confirming the picture of an O(3) model with a one--particle spectrum 
(see the discussion in Section~2).

Table~II  shows also a test of asymptotic scaling, reporting the quantity 
$(\xi_{G} f(\be)_{2l})^{-1}$, where $f(\be)_{2l} = 2\pi\be e^{-2\pi\be}$ is 
the 2--loop renormalization group function. In the asymptotic scaling region 
this quantity should be constant. However, numerical outcomes exhibit a 
steady growth and, moreover, they are far from the theoretical value
$M_G/\Lambda_{FP}$ = 9.802494 ($\Lambda_{FP}$ is the $\Lambda$ parameter of 
the theory regularized with the FP action) coming from the exact 
result for $M_G$ of Ref.~\cite{HMN90} and from the relation $\Lambda_{FP} = 
8.17\ \Lambda_{ST}$~\cite{HN94}. We stress that this is not in contradiction 
with the properties of the FP action which do manifest 
only in the physical scaling of the lattice theory. Our hypothesis is that 
the deviation from asymptotic scaling is due to large non--universal 
corrections in the function $f(\be)$. We have fitted data in the seventh 
column of Table~II taking into account the first non--universal correction: 
$f_{3l}(\be) = f(\be)_{2l}( 1 +\delta/(2\pi\be) )$. Imposing for 
$M_G/\Lambda_{FP}$ the previously quoted theoretical value, we find $\delta =
- 1.746(23)$, $\chi^2$/d.o.f.~$\simeq$~0.8. The higher order corrections to 
$f(\be)_{2l}$ are non--negligible in the region of $\be$ accessible to Monte 
Carlo simulations and it is essential to include them when the scaling
of physical quantities is studied with this kind of action.
Involving in the fit procedure also the ratio $M_G/\Lambda_{FP}$, we
find   $M_G/\Lambda_{FP}$ = 9.97(5), $\delta = - 1.80(3) $, 
$\chi^2$/d.o.f. $\simeq$ 0.9.
Nevertheless, we observe that the uncertainty of this 
measure is much larger than the na\"{\i}ve error of the fit here quoted.
A more reliable error can be found by checking the stability of the
result  under various alternative fits. Our conclusion is that
the real uncertainty is about 10\%, being therefore our determination of 
$M_G/\Lambda_{FP}$ compatible with the prevision of Ref.~\cite{HMN90}.

\section{\bf Stability of lattice classical configurations}
\label{sec:stability}

When instantons are discretized on the lattice, generally their action is 
no more scale invariant. In the case of the standard lattice action,
for instance, the action of instantons decreases with their size~\cite{Lue82}
and, as a consequence, instantons are not stable under local minimization of 
the action, but shrink up to destruction~\cite{Ber81}.

The FP action is scale invariant~\cite{HN94} (up to a minimum size 
\cite{BBHN94}) and so its topological classical solutions are stable under 
minimization. When a parametrized form of the FP action is used in numerical 
simulations the statement about stability of instantons becomes likely 
approximate: however, a huge improvement is expected with respect to the 
standard action, and with respect to the tree level Symanzik improved action
\cite{Sym80} too, where only ${\cal O}(a^2)$ cut--off effects are removed.

Anyway, we did not study the absolute stability of instantons under 
minimization, our main interest being to check to what extent a fixed 
(finite) number of steps of local minimization modifies the lattice 
topological structures. This is a way to have control over the loss of 
topological signal which affects the lattice determination of the topological
susceptibility by cooling techniques~\cite{Ber81}\cite{Tep86} 
(see Section~7).

In view of this, we have discretized on a $60^2$ lattice the exact 
2--instanton continuum solution defined on a torus~\cite{RR83}\footnote{We 
thank M. Blatter and R. Burkhalter for many helpful suggestions in this 
regard.} (1--instantons do not exist on a torus~\cite{RR83}\cite{MS94}). 
The study consisted in the observation of the  behavior during 100 steps of 
cooling of both the topological charge and  the action of the discretized  
2--instanton. We studied 2--instantons of size\footnote{The size is defined 
as in the case of instantons on the sphere~\cite{WKB}.} ranging from $0.2a$ 
to $5a$, comparing the performances of the FP and tree level Symanzik 
improved actions.
We have adopted an improved version of the lattice charge operator 
(see Section~7). The local minimization of the lattice action was obtained 
by making use of the Metropolis algorithm in the usual fashion, but with the 
request that only upgradings which lower the action are accepted.
A sweep of this $\be = \infty$ Metropolis is a cooling step.
As a preliminary task we have checked that the efficiency of the algorithm 
of minimization is independent of the action; this
test has been realized by cooling lattice configurations with trivial 
topology, and verifying that the behavior of charge and action is similar
in the two cases. 

The results of the study on the 2--instanton are summarized in Figs.~1 and 2.
We observe that in the case of the FP action topological charge loss is tiny 
if the size of the 2--instanton is larger than $0.9a$. For size smaller than 
$0.9a$ the discretized configuration becomes unstable and it is destroyed by
few cooling steps. The deviation from 2 of the topological charge of the 
cooled instantons with size smaller than $3a$ is explained by a residual 
scale dependence of our (only partially improved) definition of the lattice 
charge operator. In the case of the Symanzik action qualitatively the same 
behavior is observed, with the important difference that the topological 
signal is almost totally lost already at size $1.5a$. 

A detailed analysis of the behavior of the charge during the cooling
procedure gives some indications about shrinking effects. Shrinking can be
observed in both the Symanzik and parametrized FP actions, increasing 
at small sizes (where the scale symmetry breaking of the 
discretization is amplified); however, it is much reduced in the case of the
FP action (see Table~III): at $\rho = 2a$, for instance, it is $\sim 2$ 
orders of magnitude smaller.

In Fig.~3 we compare the different cooling shapes of a small--size 2--instanton
($\rho = 1.2a$) when the FP, the standard and the Symanzik actions are used; 
being the algorithm of minimization non--deterministic, we have checked to what
extent results fluctuate under changing of the sequences of random numbers used
in the minimization algorithm.
 
Although single--instanton classical solutions are not allowed on a torus,
it is nevertheless interesting to investigate the amount of topological loss 
under cooling in the $Q=1$ topological charge sector, which is dominant 
among classical configurations at thermal equilibrium. It is impossible to 
perform a study strictly equivalent to the previous one in the Q=2 sector, 
because 1--instanton solutions are missing on a torus: in their place, we
have used 1--instanton solutions with constant boundary conditions 
\cite{BP75} which are quasi--solutions for $\rho/L \ll 1$ 
($L$ is the lattice size). We have found substantially the same results of 
Figs.~1 and 2 as far as the minimum sizes are concerned: the main 
difference is that the observed  action is slightly above the expected 
value $4\pi$ at large $\rho$, going asymptotically to $4\pi$ when 
$\rho/L$ decreases (in Fig.~4 we report the results for the FP action); 
clearly this is a boundary effect (the configuration does not satisfy the 
correct periodic boundary conditions).

\section{\bf Topology: the geometrical method}
\label{sec:geo}

In this Section we check how the geometrical method for the determination
of the topological susceptibility~\cite{BL81} works when the FP action is 
used.

The geometrical definition $Q_g$ of the lattice topological charge is an 
attempt to recover the topological structure of the continuum theory by
interpolating the discrete lattice field with a continuous field on which
the topological charge can be measured.
In formulae:
\beq
Q_g = \sum_{x^*} q(x^*)\;\;\;, 
\eeq
where $x^*$ indicates a vertex of the dual lattice (corresponding to
a plaquette of the 2--d lattice) and 
\beq
q(x^*) = \f{1}{4\pi} \left\{ (\sigma A) (\phi_1,\phi_2,\phi_3) + 
		           (\sigma A) (\phi_1,\phi_3,\phi_4) \right\}\;\;\; .
\eeq
$\phi_1$, $\phi_2$, $\phi_3$, $\phi_4$ are the four spins belonging to the 
plaquette $x^*$ and $(\sigma A)(\phi_1,\phi_2,\phi_3)$ denotes the 
signed area of the spherical triangle  $\phi_1$, $\phi_2$, $\phi_3$ 
\footnote{One can demonstrate  that
\begin{eqnarray}
\exp\left(\f{1}{2}i\sigma A\right) = \rho^{-1}
\left\{1+\phi_1\cdot\phi_2+\phi_2\cdot\phi_3+\phi_3\cdot\phi_1+i\phi_1\cdot\left(\phi_2\times\phi_3\right)\right\}\;\;\; , \nonumber \\
\rho=\left\{2\left(1+\phi_1\cdot\phi_2\right)\left(1+\phi_2\cdot\phi_3\right)\left(1+\phi_3\cdot\phi_1\right)\right\}^{1/2}>0\;\;\;  ,
\label{eq:geodef}
\end{eqnarray}
thus providing an explicit formula for $Q_g\ $.}.

The topological susceptibility constructed in terms of the geometrical
charge $\chi_g = \langle Q_g^2/V \rangle$~\cite{BL81} fails to reproduce the 
correct continuum behavior: the reason is that the geometrical definition 
assigns non--zero topological charge to lattice structures of unphysical size
(dislocations). In the case of the standard lattice action, dislocations 
proliferate in the $\be\rightarrow\infty$ limit~\cite{Lue82}\cite{Ber81} 
since their action is definitely lower than $4\pi$, the minimal (physical) 
value in the $Q=1$ sector. As a consequence, they produce a non--scaling 
topological signal.
Now, it is interesting to investigate to what extent the situation improves 
with the FP action, checking, in particular, if dislocations are still 
present and, if so, if their action is at least closer to $4\pi$. In order to
have an indication about this subject, we have studied the cooling of a 
discretized 1--instanton solution (with constant boundary solutions to 
infinity). This lattice configuration -- which is not a local minimum
of the action on a toroidal lattice -- is driven by the cooling algorithm
to the nearest minimum in the $Q_g=1$ sector (lying on the border between the
$Q_g=1$ to the $Q_g=0$ sectors), and eventually it passes into 
the $Q_g=0$ sector. The action of this border configuration is, in the case
at hand (see Fig.~5), $S/4\pi=0.92 < 1$, thus indicating a dislocation. This 
outcome (though the method was not rigorous) suggests that the 
action of dislocations is closer to $4\pi$ compared to the case of the 
standard action, where the minimal action in the $Q_g=1$ sector is 
0.53~\cite{Lue82}. Anyway, dislocations are still present, so, apart from
non--scaling effects due to the UV topological dominance of the 
model~\cite{MS94}\cite{FP94}, a deviation of $\chi_g$ from the renormalization 
group behavior is expected.

We have performed numerical simulations 
on $\chi_g$ (see  Fig.~6 and Table~IV) finding $\chi_g \sim 
\be^2 e^{-4\pi\alpha\be}$, with $\alpha=0.8315(11)$, $\chi^2/$d.o.f. 
$\simeq$ 0.8. For comparison, we report the $\alpha$ coefficient
obtained in the case of the standard action~\cite{BL81}, $\alpha=0.625$: 
clearly  the effect of dislocations is lower with the FP action.

\section{\bf Topology: the field theoretical method}
\label{section-ftm}

\subsection{\bf Perturbative calculations}

We follow the field theoretical method~\cite{FTM} for the determination of 
the topological susceptibility from the lattice. First of all,
a lattice topological charge density operator is defined
as a local operator having the appropriate classical continuum limit
\cite{DKRV81}; our choice is~\cite{DFPV92}:
\beq
Q^L(x) \: = \:\frac{1}{32\pi } \epsilon _{\mu \nu }\epsilon _{ijk} \phi _{i}(x)
\left(\phi _{j}(x+\mu ) \: - \: \phi _{j}(x-\mu ))(\phi _{k}(x+\nu ) \: - \: 
\phi _{k}(x-\nu )\right) \;\;\; .
\label{eq:latch}
\eeq

From the prescriptions of field theory it comes that 
a finite multiplicative renormalization connects
the matrix elements of the lattice topological charge density with those
of its continuum counterpart:
\beq
Q^{L}(x) \: = \: a^2 \, Z(\beta) \: Q(x) \: + \: O(a^4) \; \; .
\label{eq:zq}
\eeq
We recall: $\be$ = $1/g$.
The lattice--regularized version of $\chi$ is 
\beq
\chi ^{L} \: = \: \left\langle \;\sum_{x} Q^{L}(x)Q^{L}(0)\; 
\right\rangle \: = \: \frac{1}{L^2}\left\langle ( \:\sum_{x} Q^{L}(x) 
\:)^{2} \right\rangle \: \:\; ,
\label{eq:susc}
\eeq
where $L$ is the lattice size.

A prescription equivalent to 
Eq.~(\ref{eq:crew}) does not exist on the lattice and therefore the
contribution of the contact terms must be isolated and subtracted. These
contact terms appear as mixings with the action density $S(x)$ and with
the unity operator $I$, which are the only available operators with 
equal dimension or lower. In formulae
\beq
\chi ^{L}(\be ) \: = \: a^{2}\:Z(\be )^{2}\,\chi \: + \: a^{2}\:A(\be )\,
\langle S(x) \rangle \: + \: P(\be ) \,\langle I \rangle \: + \: O(a^{4}) \: 
\: .
\label{eq:ope}
\eeq 
In Eq.~(\ref{eq:ope}) the quantity
$\langle S(x) \rangle$ is intended to be the non--perturbative part of
the expectation value of the action density, i.e. it is a signal of
dimension two.

The extraction of the physical value of the topological susceptibility
from numerical simulations requires the evaluation of the renormalization
constants $Z(\be)$, $A(\be)$, $P(\be)$.

The standard way to proceed is to perform the calculation in perturbation
theory. We have decided to consider in our calculations the quartic terms
of the parametrization of the FP action  with coefficient 
${\cal O}(10^{-2})$ or larger. The procedure 
followed for the calculation of $Z(\be)$ and $P(\be)$ is  illustrated in
Ref.~\cite{DFPV92}; for $A(\be)$ we refer to Ref.~\cite{FP94}.

An intermediate step in the calculation of $Z(\be)$ is the evaluation of
the renormalization constants of the fields $\pi(x)$ and of the
coupling constant $g$. We have found them to 1--loop order by imposing
\beq
Z_\pi^{\overline{MS}}(g,\mu a)\Gamma_{\pi\pi}^{L}(g,h,a;p) \, = \, 
\Gamma_{\pi\pi}^{\overline{MS}}(g_r,h_r,\mu ;p) \;\; , 
\label{eq:gamma2}
\eeq
where $g = Z_g^{\overline{MS}} g_r$, $h =  Z_g^{\overline{MS}} 
{Z_{\pi}^{\overline{MS}}}^{-1/2} h_r$. We quote here the results:
\bea
Z_{\pi}^{\overline{MS}}(g,\mu a) & = & 1 - 2L(\mu a)g + O(g^2) \;\; , 
\\ \nonumber
Z_g^{\overline{MS}}(g,\mu a) & = & 1 - ( L(\mu a) + c_1)g + O(g^2) \;\; ,
\eea
where $L(x) = -\frac{1}{2\pi} \ln x + \frac{5}{4\pi}\ln 2 - y$, being 
$y = 0.0382$ and $c_1 = 0.1262$. We observe that at this order 
$Z_\pi^{\overline{MS}}$ is independent from the couplings of the multi--spin 
terms  which affect only  $c_1$.

For the calculation of $Z(\be)$ we compare the 2--point $\Gamma$--functions 
with one operator insertion on the lattice and on the continuum:
\beq
Z(\be)^{-1}Z_\pi^{\overline{MS}}(g,\mu a)\Gamma_{Q,\pi\pi}^{L}(g,h,a;p,q)\, 
= \, \Gamma_{Q,\pi\pi}^{\overline{MS}}(g_r,h_r,\mu ;p,q) \;\; . 
\label{eq:gamma2q}
\eeq
$\Gamma_{Q,\pi\pi}$ does not take contribution from the couplings 
of multi--spin terms at 1--loop order, so our 1--loop  determination 
of $Z(\be)$ is exact.
We show the result:
\beq
Z(\be) \; = \; 1 + \f{z_1}{\be} + O(\be^{-2})\;\; , 
\;\;\;\; z_1 = - 0.94237 \;\; .
\eeq

We calculated the perturbative tail $P(\be) = \sum_{n=4} p_n /\be^n$ 
up to four loops. The results for infinite volume are:
\beq
p_4 \; = \; 1.97429 \times 10^{-4}\;\; , \;\;\;\; 
p_5 \; = \; -2.24879 \times 10^{-4}\;\; .
\eeq
Only $p_5$ is affected by our approximation. These numbers were extracted by
performing numerical integrations at finite volume and extrapolating the 
results to infinite volume. 

We have calculated the $a_3$ coefficient in the perturbative series 
$A(\be) =\sum_{n=3} a_n/\be^n$ by comparing
$\Gamma_{(Q^L)^2/V,\pi\pi}$ with $\Gamma_{S,\pi\pi}$ at the lowest order
(for details see the Appendix in Ref.~\cite{FP94}). The value of $a_3$
extrapolated to infinite volume is $7.76206 \times 10^{-4}$. This
result also is independent from the
multi--spin terms. As in the case of the tree level Symanzik
improved action~\cite{FP94}, the mixing
of the topological susceptibility with the action density is absolutely
irrelevant.

\subsection{\bf Numerical simulations}

We have performed Monte Carlo simulations for $\chi^L$ on a $60^2$ lattice
over an extended region of values of $\be$, the algorithm being the same
4--hits Metropolis of previous simulations. In Table~V we report the results:
binning techniques have been applied in order to take account of the effect 
of correlations in the evaluation of errors. In order to control finite size 
effects, we have repeated the simulations at $\be = 1.0, 1.1, 1.2$ with the 
care that $L/\xi_G \sim 6$. We observe (see Table~V) that only the datum at 
$\be = 1.2$ appreciably differs from the result on the $60^2$ lattice.

In the range of values of $\be$ accessible to our lattices, the perturbative 
determinations of the previous subsection are not sufficient to have a 
reliable estimate of the renormalizations $Z(\be)$ and $P(\be)$. Moreover, 
the standard technique to estimate some further terms in the perturbative 
expansion of  $P(\be)$ by a fit on numerical data at $\be$ higher than a 
certain $\be_t$ (the so--called perturbative region) does not work:
the extrapolation of $P(\be)$ to  lower values of $\be$, where the physical 
signal is still living, is not reliable, since this procedure is unstable 
under  changes of the conditions of the fit.
So, we followed an alternative method for the determination of the 
renormalizations of the lattice topological susceptibility: the ``heating" 
method~\cite{Tep89}\cite{DV92}. The heating method, relying only on MC 
techniques, is fully non--perturbative. It consists (for details see also
\cite{DFPV92}\cite{FP94}) in constructing on the lattice ensembles of 
configurations $\{C_t\}$, each configuration of the ensemble being obtained 
by performing a sequence of $t$  local Monte Carlo sweeps starting from
a discretized classical configuration $C_0$ -- a large instanton or 
the flat configuration. For small $t$, the heating process thermalizes only
the small--range fluctuations which are responsible for the 
renormalizations: when the starting configuration is a large instanton
(flat configuration), measuring $Q^{L}$($\chi^L$) on the ensembles 
$\{C_t\}$, a plateau at the value of  $Z(\be)$ $\,(P(\be))$ is observed after
a certain time, not depending on $\be$, corresponding to the 
thermalization of quantum fluctuations.

In Table~VI we show the values of  $Z(\be)$ evaluated by the heating 
procedure at various $\be$. Data reported therein have been obtained by 
heating a discretized 1--instanton with constant boundary 
conditions\footnote{For the purposes of this method it
is sufficient to thermalize around any smooth configuration with
non--trivial topology.} (size 10 lattice units) on a 
$60^2$ lattice. The  ensembles $\{C_t\}$ have been obtained performing $t$ 
sweeps of the Metropolis algorithm. 
In Table~VI we also compare the non--perturbative determinations
of $Z(\be)$ with our 1--loop perturbative calculation: the latter well 
reproduces the non--perturbative results at the largest values of $\be$, but 
a deviation is observed at the smallest ones, where the 1--loop perturbative 
expansion is expected to fail. Attributing  this discrepancy to the 
next--to--leading terms of the perturbative expansion of $Z(\be)$, we have 
found by a fit on data for $\be\leq1.4$:  $z_2 = 0.473(7)$, $z_3 = 
-0.234(9)$ ($\chi^2/$d.o.f. $\simeq$ 0.1).

In Table~VII we show the outcomes of  the heating method for $P(\be)$ at 
various $\be$. Data have been obtained on a $60^2$ lattice. Using data from 
heating, we have fitted two further coefficients, $p_6$ and $p_7$, of the 
perturbative expansion of $P(\beta)$; the result is: $p_6 = (7.54 \pm 
0.09)\times 10^{-5}$, $p_7 = (3.02 \pm 0.10)\times 10^{-5} $, with 
$\chi^2/$d.o.f. $\simeq$ 0.9. In Fig.~7 we compare three different 
determinations of $P(\be)$: the first is the pure 4--loop perturbative 
calculation; the second is a fit which combines the perturbative calculation 
with the equilibrium values of $\chi^L$ at values of $\be$ larger than 1.6; 
the last is the non--perturbative determination of $P(\be)$. As it can be 
clearly observed, the curve from the second determination badly reproduces 
data from the heating method for the values of $\be$ not included in the fit,
i.e. $\be < 1.6$. This explains why the standard procedure of extracting the 
perturbative tail joining perturbative calculations and MC data from the 
perturbative region fails in this case.

Having obtained a reliable determination of the renormalizations
$Z(\be)$ and $P(\be)$, we are now in a position to  extract from 
$\chi^L$ the physical quantity $\chi$ (see Eq.~(\ref{eq:ope}) --  
the term proportional to $A(\be)$ is negligible).

In Table~VIII (see also Fig.~8) we report, for different values of $\be$, 
the quantities $\chi/\Lambda_{FP}^2$ and $\chi/M_G^2$ (in the latter entry we
use the relation between $\Lambda_{FP}$ and $M_G$ which
can be deduced  from  Ref.~\cite{HMN90}); we also report 
our determination of $\xi_G$, as obtained from the fit of Section~3
(the one with the ratio $M_G/\Lambda_{FP}$ fixed at the theoretical value of 
Ref.~\cite{HMN90}). 

In Table~IX we compare our present results for  $\chi/M_G^2$ with other 
determinations known in literature. Data in columns (2) and (3) have been
obtained in  the field theoretical approach with the standard definition
of the topological charge operator, but using other lattice regularizations 
of the action~\cite{DFPV92}\cite{CRV92}. In column (4) a parametrized 
FP action and the FP lattice charge operator have been used. The FP charge 
operator has been constructed by the application of one RG transformation on 
the geometrical charge operator~\cite{BBHN94}. A substantial agreement is
observed among determinations using the standard charge operator in the field
theoretical approach, while data for $\chi/M_G^2$ of column (4), 
where a FP charge operator is used, are systematically larger.
We argue that the disagreement originates from the different performance in 
the short distance regime of the two lattice definitions of the charge 
operator. Indeed, the field theoretical definition is likely to
underestimate the  topological content of small size topological
configurations (as Fig.~1 clearly singles out), while such loss of 
topological signal is definitely less relevant with the FP operator of 
Ref.~\cite{BBHN94} (Fig.~2 of the previously quoted
reference). This different behavior strongly reflects upon the determination 
of the topological susceptibility, owing to the UV topological dominance of 
the model.

\section{\bf Topology: the cooling method}
\label{sec:cooling}

In this Section we turn to an alternative way to face the matter of 
renormalizations, known in literature as the cooling method 
\cite{Ber81}\cite{Tep86}. The method relies on  a local algorithm of  
minimization of the action of lattice configurations -- the cooling; the 
purpose is to destroy the quantum fluctuations of the lattice configuration 
generated by the MC algorithm (which give rise to the renormalizations
$Z(\be)$ and  $P(\be)$ of Eq.~(\ref{eq:ope})), trying to preserve the 
background topological structure. 
If the background structure is an extended classical configuration, after 
a large enough number of cooling iterations, a measure of the lattice 
topological charge on the cooled configuration well approximates an 
integer\footnote{If the background configuration is not slowly varying, scale
effects produce deviations from integer values, due to the scale dependence
of the field theoretical charge operator (see Section~4).}.

As we have seen in Section~4, shrinking effects are tiny with the 
(parametrized) FP action, so the most important drawback of the cooling 
procedure is overcome.

In Table~V we report the results of the simulations on a $60^2$ lattice.
The minimization  algorithm has been described in  Section~4;
the ``cooled'' equilibrium value of the topological susceptibility
has been determined  by measuring the topological charge on configurations 
obtained from the equilibrium ones after 30 iterations of the
algorithm.

Here we adopt a different definition of  the lattice charge operator.
It consists in a tree level Symanzik improved operator, where the
first three irrelevant terms have been eliminated:
\beq
Q^{L}(x) \: = \: \frac{1}{8\pi } \epsilon _{\mu \nu }\epsilon _{ijk} 
\phi _{i}(x) D_{\mu}\phi _{j}(x)D_{\nu}\phi _{k}(x)\;\;\;;
\eeq
$D_{\mu}$ is the improved lattice derivative:
\bea
&D_{\mu}\: = \:\f{1225}{1024}\nabla^{(1)}_{\mu}-
\f{245}{1024}\nabla^{(3)}_{\mu}+\f{49}{1024}\nabla^{(5)}_{\mu}
-\f{5}{1024}\nabla^{(7)}_{\mu} \;\;\;,&\\
&\nabla^{(n)}_{\mu}\phi_{i}(x)\: =\:
\f{1}{2n}(\phi _{i}(x+n\mu ) - \phi _{i}(x-n\mu )) \;\;\;.& \nonumber
\eea
Being the first irrelevant terms absent, this operator is less
sensitive to quantum fluctuations than the definition of Eq.~(\ref{eq:latch})
and, as a consequence, the determination of the topological charge
of the cooled configurations improves.

For each value of $\be$ typically 100000 configurations were generated, 
the cooling procedure being performed once every 100 configurations. 
In Table~VIII and in Fig.~8 we compare the cooling determinations
with the results of the field theoretical method.
Data from cooling nicely reproduce the outcomes of the field
theoretical method, so confirming that topological structures are affected
in a negligible way by the cooling procedure.

The results for $\chi/M_G^2$ obtained on large lattices by cooling (see 
Table~VIII) allow now to appreciate a slight deviation from scaling -- a 
non--scaling behavior is observed also in Ref.~\cite{BBHN94}. Such 
deviation is expected since a ($\be$--dependent) part of the whole topological 
signal is lost on the lattice due to the UV dominance of the model.  
However, the discrepancy of our data with the results of the previously quoted 
reference is still present and so it cannot be explained by finite size
effects.

\vspace{1truecm}
{\bf Acknowledgments.} We wish to thank Adriano Di Giacomo for having 
suggested the subject, M.~Blatter, R.~Burkhalter, A.~Hasenfratz, 
P.~Hasenfratz, F.~Niedermayer, P.~Rossi and E.~Vicari for many useful 
discussions. This work was partially supported by  Fondazione
``A. Della Riccia'' (Italy).

\newpage

\noindent
TABLE I: correlation length $\xi$ and correlation function coefficient
$A$ with the FP action.
\newline

\begin{tabular}{||c|c|c|c|c|c|c|c||} \cline{1 - 8}  
$\;\;\;\beta\;\;\;$ &$\;\;\; L\;\;\;$ & $\;\;\;\xi_{G}\;\;\;$ & $\;\;\;\xi_{w}\;\;\;$ & $\;\;\;\xi_{d}\;\;\;$ & $\;\;\;A_G\;\;\;$ & $\;\;\;A_w\;\;\;$ & 
$\;\;\;A_d\;\;\;$   \\
\cline{1 - 8}  
0.6 & 36 & 2.14(4) & 2.188(15) & 2.180(21) & 5.15(18) & 4.84(4)  & 4.84(6) \\
0.7 & 36 & 3.14(5) & 3.16(5)   & 3.17(5)   & 6.3(1)   & 6.22(11) & 6.19(12)\\
0.75& 36 & 3.89(7) & 3.89(7)   & 3.89(6)   & 7.04(7)  & 6.96(12) & 7.00(9) \\
0.8 & 36 & 4.81(7) & 4.83(9)   & 4.80(6)   & 8.103(23)& 8.01(14) & 8.11(8) \\
0.85& 36 & 5.87(9) & 5.86(13)  & 5.97(9)   & 9.37(3)  & 9.45(17) & 9.30(10)\\
0.9 & 48 & 7.34(16)& 7.34(21)  & 7.34(15)  & 10.98(3) & 10.99(25)&11.12(14)\\
1.0 & 48 &11.71(29)&11.8(4)    &11.89(27)  & 15.01(14)& 15.0(3)  &15.4(3)  \\
1.1 &130 &18.5(2.1)& 18.1(2.7) &19.1(2.1)  & 22.3(4)  & 24(4)    &22.2(1.7)\\
1.15&150 &24.4(1.9)& 24.3(2.5) &23.7(1.6)  & 26.5(3)  & 26.7(2.5)&28.2(1.2)\\
1.2 &150 & 34(3)   & 34(4)     &33.0(2.9)  & 33.1(1.3)& 32(3)    &35.1(2.6)\\
\cline{1 - 8}  
\end{tabular}
\vspace{1.5truecm}

\noindent
TABLE II: ratios of different definitions of correlation length and
correlation function coefficient; in the last column we show 
$(\xi_{G} f(\be)_{2l})^{-1}$. 
\newline

\begin{tabular}{||c|c|c|c|c|c|c||} \cline{1 - 7}  
$\;\;\;\beta\;\;\;$ &$\;\;\; L\;\;\;$ &$\;\;\;\xi_{G}/\xi_w\;\;\;$ & 
$\;\;\;\;\xi_d/\xi_w\;\;\;\;$ & $\;\;\;A_w/A_G\;\;\;$ & $\;\;\;A_d/A_w\;\;\;$
 & $(\xi_{G} f(\be)_{2l})^{-1}$  \\
\cline{1 - 7}  
0.6  &  36 & 0.976(18)  & 0.997(7)  & 0.94(3)   & 1.001(11) & 5.59(23)  \\
0.7  &  36 & 0.994(5)   & 1.003(9)  & 0.987(11) & 0.996(17) & 5.88(10)  \\
0.75 &  36 & 0.9977(21) & 0.998(11) & 0.988(9)  & 1.006(18) & 6.08(11)  \\
0.8  &  36 & 0.997(4)   & 0.994(11) & 0.989(14) & 1.011(19) & 6.28(9)   \\
0.85 &  36 & 1.002(6)   & 1.019(15) & 1.009(19) & 0.984(22) & 6.66(11)  \\
0.9  &  48 & 1.000(6)   & 0.999(18) & 1.001(22) & 1.012(27) & 6.88(16)  \\
1.0  &  48 & 0.993(12)  & 1.008(25) & 1.00(3)   & 1.03(3)   & 7.28(18)  \\
1.1  & 130 & 1.02(4)    & 1.06(11)  & 1.09(15)  & 0.91(15)  & 7.8(9)    \\
1.15 & 150 & 1.00(3)    & 0.98(6)   & 1.01(10)  & 1.05(10)  & 7.8(6)    \\
1.2  & 150 & 0.99(5)    & 0.96(10)  & 0.97(13)  & 1.09(16)  & 7.5(6)    \\
\cline{1 - 7}  
\end{tabular}

\newpage

\noindent
TABLE III: topological charge (in the field theoretical definition) of 
2--instantons of different initial size at the beginning and after 100 step 
of the cooling procedure with the Symanzik and the FP actions. The lattice
size is 60.
\newline

\begin{center}
\begin{tabular}{||c|c|c|c||} \cline{1 - 4}
          & $Q_0$ & \multicolumn{2}{c||}{$Q_{100}$} \\ \cline{ 3 - 4}
 $\rho/a$ &                &  Symanzik action  & \hspace{.15in}  FP action 
\hspace{.15in} \\  \cline{1 - 4}
 4  &   1.99856751 &     1.99856600  &          1.99856713  \\ 
 2  &   1.93896373 &     1.92357210  &          1.93858846  \\
1.2 &   1.64901344 &     0.00001479  &          1.64991933  \\ \cline{1 - 4}
\end{tabular}
\end{center}

\vspace{1.5truecm}

\noindent
TABLE IV: $\chi_g$, $\chi_g/\Lambda_{FP}^2$ and $\chi_g/M_G^2$ for various
$\be$ and  $\xi_G$ on a $60^2$ lattice. 
$\xi_G$ is obtained from the fit of Section~3
(the one with the ratio $M_G/\Lambda_{FP}$ fixed at the theoretical value of 
Ref.~\cite{HMN90}).
\newline

\begin{center}
\begin{tabular}{c||c|c|c|c|c||c} \cline{2 - 6}
&$\;\;\;\beta\;\;\;$ & $\;\;\;\;\;\;\;\xi_G\;\;\;\;\;\;\;$ & 
$ \;\;\;\; \chi _g\times 10^4 \;\;\;\; $ &
$\;\;\chi_g/\Lambda_{FP}^2\;\;$  & $\;\;\;\;\chi_g/M_G^2\;\;\;\;$ &
 \\  \cline{2 - 6}
& 1.05  & 15.42(7)   &  8.27(26)  & 18.9(8)   & 0.197(8)  & \\
& 1.1   & 19.82(9)   &  5.63(21)  & 21.2(1.0) & 0.221(10) & \\
& 1.15  & 25.58(11)  &  3.61(9)   & 22.7(8)   & 0.236(8)  & \\
& 1.2   & 33.13(13)  &  2.25(7)   & 23.7(9)   & 0.247(10) & \\
\cline{2 - 6}
\end{tabular}
\end{center}

\newpage

\noindent 
TABLE V: $\chi^L$ and $\chi_{cool}$ versus $\be$ on a $60^2$ lattice.
Data with $^*$, $^{**}$, and $^{***}$ are obtained on lattices $70^2$, 
$130^2$ and $180^2$, respectively.
\newline

\begin{center}
\begin{tabular}{c||c|c|c||c|c||c} \cline{2 - 6}
&$ \beta $ & $\;\;\;\chi^L\times 10^5\;\;\;$ & $\;\chi_{cool}\times 10^4\;$ & 
$\;\;\;\; \beta\;\;\;\; $ & $\;\;\;\chi^L\times 10^5\;\;\;$ & 
\\  \cline{2 - 6}
& 0.8          & 27.0(4)    &   -      & 1.5     & 1.91(6)    &   \\
& 0.85         & 23.8(3)    &   -      & 1.52    & 1.94(9)    &   \\
& 0.9          & 20.30(29)  &   -      & 1.55    & 1.77(8)    &   \\
& 0.95         & 17.13(18)  & 9.2(5)   & 1.58    & 1.61(7)    &   \\ 
& 1.           & 14.12(16)  & 6.42(29) & 1.6     & 1.54(5)    &   \\
& $^*1.$       & 13.97(20)  & 5.98(27) & 1.65    & 1.39(6)    &   \\
& 1.05         & 11.28(12)  & 4.48(21) & 1.7     & 1.23(5)    &   \\ 
& 1.1          &  9.18(10)  & 2.79(19) & 1.75    & 1.06(6)    &   \\
& $^{**}1.1$   &  9.48(25)  & 2.9(3)   & 1.8     & 0.98(5)    &   \\
& 1.12         &  8.01(9)   & 2.35(13) & 2.      & 0.68(3)    &   \\ 
& 1.14         &  7.36(9)   & 1.88(11) & 2.2     & 0.487(15)  &   \\
& 1.16         &  6.72(8)   & 1.48(9)  & 2.3     & 0.426(14)  &   \\ 
& 1.18         &  6.05(8)   & 1.33(10) & 2.5     & 0.326(10)  &   \\
& 1.2          &  5.42(6)   & 1.00(9)  & 2.8     & 0.199(6))  &   \\ 
& $^{***}1.2$  &  6.75(25)  & 1.27(13) & 3.      & 0.160(5)   &   \\
& 1.22         &  5.3(3)    &   -      & 3.5     & 0.095(3)   &   \\
& 1.225        &  4.91(22)  &   -      & 4.      & 0.0576(9)  &   \\ 
& 1.25         &  4.50(17)  &   -      & 4.5     & 0.0384(10) &   \\
& 1.3          &  3.82(15)  &   -      & 5.      & 0.0256(7)  &   \\ 
& 1.32         &  3.35(10)  &   -      & 6.      & 0.0131(9)  &   \\
& 1.35         &  3.06(15)  &   -      & 7.      & 0.00683(18)&   \\ 
& 1.38         &  2.96(15)  &   -      & 8.      & 0.00409(11)&   \\
& 1.4          &  2.72(23)  &   -      & 9.      & 0.00250(7) &   \\ 
& 1.42         &  2.40(7)   &   -      &10.      & 0.00178(5) &   \\
& 1.45         &  2.36(18)  &   -      &         &            &   \\ 
\cline{2 - 6}
\end{tabular}
\end{center}

\newpage

\noindent
TABLE VI: $Z(\beta)$ versus $\beta$. $Z(\beta)_{1l}$ is the 
multiplicative renormalization calculated to one loop.
$Z(\beta)_{MC}$ is the multiplicative renormalization
calculated by heating a discretized $Q=1$ smooth configuration; the size of 
the lattice is 90 for $\be\geq$ 1.5 and 60 for the remainder;
the statistic of the simulations is 1000. Since data on the plateau are
correlated, we have estimated  errors by jackknife techniques.
\newline

\begin{center}
\begin{tabular}{c||c|c|c||c} \cline{2 - 4}
& $\;\;\;\;\;\beta\;\;\;\;\;$ & 
$\;\;\;\;\; Z(\beta)_{1l}\;\;\;\; $  & $\;\;\;\;Z(\beta)_{MC}\;\;\;\; $ & 
\\  \cline{2 - 4}
&  5.0       &  0.812                   &  0.8247(7)       &  \\
&  4.5       &  0.791                   &  0.8096(8)       &  \\
&  4.0       &  0.746                   &  0.7865(11)      &  \\
&  3.5       &  0.731                   &  0.7599(13)      &  \\
&  3.0       &  0.686                   &  0.7254(18)      &  \\         
&  2.5       &  0.623                   &  0.6804(25)      &  \\
&  2.1       &  0.551                   &  0.631(3)        &  \\
&  1.9       &  0.504                   &  0.591(4)        &  \\
&  1.7       &  0.446                   &  0.566(4)        &  \\
&  1.5       &  0.372                   &  0.514(6)        &  \\
&  1.4       &  0.327                   &  0.479(9)        &  \\
&  1.35      &  0.302                   &  0.466(10)       &  \\
&  1.3       &  0.275                   &  0.450(10)       &  \\
&  1.25      &  0.246                   &  0.430(11)       &  \\
&  1.2       &  0.215                   &  0.406(12)       &  \\
&  1.15      &  0.181                   &  0.380(13)       &  \\
&  1.1       &  0.143                   &  0.362(14)       &  \\
\cline{2 - 4}
\end{tabular}
\vspace{1.5truecm}
\end{center}

\newpage

\noindent
TABLE VII: $P(\beta)_{np}$ versus $\beta$. $P(\beta)_{np}$ is 
 the mixing with the unity operator $I$ calculated by heating the 
flat configuration; the size of the lattice is 60, the statistic of 
the simulation is 1000.
\newline

\begin{center}
\begin{tabular}{||c|c||c|c||}    \cline{1 - 4}
 $\;\;\;\;\;\;\beta\;\;\;\;\;\;$ & 
$\;\;\;P(\beta)_{np}\times 10^{5}\;\;\;$ & $\;\;\;\;\;\;\beta\;\;\;\;\;\;$ & 
$\;\;\;P(\beta)_{np}\times 10^{5}\;\;\;$  
\\ \cline{1 - 4}
  1.    &    7.58(24)     &  1.25  &    3.30(8)      \\
  1.05  &    6.48(19)     &  1.3   &    2.84(6)      \\
  1.1   &    5.49(14)     &  1.35  &    2.53(6)      \\
  1.15  &    4.60(11)     &  1.4   &    2.23(5)      \\
  1.2   &    3.84(9)      &        &                 \\   
\cline{1 - 4}
\end{tabular}
\end{center}

\vspace{1.5truecm}

\noindent
TABLE VIII: $\chi/\Lambda_{FP}^2$ and $\chi/M_G^2$ for various
$\be$ and  $\xi_G$ for both the field theoretical and
the cooling determinations. $L$ is the size of the lattice.
\newline

\begin{center}
\begin{tabular}{||c|c|c|c|c|c|c||} \cline{1 - 7}
$\;\;\;\beta\;\;\;$ & $L$ & $\;\;\;\;\;\;\;\xi_G\;\;\;\;\;\;\;$ & 
$\;\;\chi/\Lambda_{FP}^2\;\;$  & $\;\;\;\;\chi/M_G^2\;\;\;\;$ &
$\chi_{cool}/\Lambda_{FP}^2$  &$\;\chi_{cool}/M_G^2\;$ \\  \cline{1 - 7}
 0.8   & 60 &  4.74(3)    & 8(7)      & 0.09(7)   &    -      &    -    \\
 0.85  & 60 &  5.92(4)    & 9(4)      & 0.09(4)   &    -      &    -    \\
 0.9   & 60 &  7.46(4)    & 8.9(2.5)  & 0.093(27) &    -      &    -    \\
 0.95  & 60 &  9.45(5)    & 9.4(1.9)  & 0.098(20) & 7.9(5)    & 0.083(5)\\
 1.    & 60 & 12.04(6)    & 10.0(1.7) & 0.104(18) & 8.9(5)    & 0.093(5)\\
 1.    & 70 &             &    -      &    -      & 8.3(5)    & 0.087(5)\\
 1.05  & 60 & 15.42(7)    & 10.3(1.5) & 0.107(16) & 10.2(6)   & 0.106(6)\\
 1.1   & 60 & 19.82(9)    & 11.3(1.5) & 0.118(15) & 10.5(8)   & 0.110(8)\\
 1.1   &130 &             &    -      &    -      & 11.1(1.3) & 0.115(13)\\
 1.12  & 60 & 21.94(10)   & 10.3(1.4) & 0.108(14) & 10.9(7)   & 0.113(7)\\
 1.14  & 60 & 24.31(10)   & 10.7(1.4) & 0.111(14) & 10.7(7)   & 0.111(7)\\
 1.16  & 60 & 26.93(11)   & 10.9(1.4) & 0.113(14) & 10.3(7)   & 0.107(7)\\
 1.18  & 60 & 29.86(12)   & 10.6(1.4) & 0.111(15) & 11.4(1.0) & 0.118(10)\\
 1.2   & 60 & 33.13(13)   & 10.0(1.3) & 0.104(14) & 10.5(1.0) & 0.110(10)\\
 1.2   &180 &             &    -      &     -     & 13.4(1.5) & 0.139(15)\\
\cline{1 - 7}
\end{tabular}
\end{center}

\newpage

\noindent
TABLE IX: Comparison of different determinations of $\chi\cdot\xi_G^2 = 
\chi/M_G^2$:
field theoretical method and  FP action (column (1), present work),  
field theoretical method and  Symanzik action (column (2), 
\cite{DFPV92}), field theoretical method and CP$^1$ standard action with 
explicit gauge degrees of freedom (column (3),~\cite{CRV92}), FP charge 
operator and parametrized FP action\footnote{We are indebted to M. Blatter 
and R. Burkhalter for the data relative to Fig.~2 in~\cite{BBHN94}.} (column
(4),~\cite{BBHN94}). 
\newline

\begin{center}
\begin{tabular}{||c|c|c|c|c||} \cline{1 - 5}
 $\;\;\;\;\;\;\;\xi_G\;\;\;\;\;\;\;$ &(1)&(2)&(3)&(4)         \\
\cline{1 - 5}
  4.74(3)   & 0.09(7)   &    -       &    -       &    -       \\
  5.92(4)   & 0.09(4)   &    -       &    -       &    -       \\
  6.057(17) &   -       &    -       &    -       & 0.1004(9)  \\
  7.46(4)   & 0.093(27) &    -       &    -       &    -       \\
  9.45(5)   & 0.098(20) &    -       &    -       &    -       \\
 12.04(6)   & 0.104(18) &    -       &    -       &    -       \\
 12.16(3)   &    -      &    -       &    -       & 0.1448(15) \\  
 15.42(7)   & 0.107(16) &    -       &    -       &    -       \\
 15.9(9)    &    -      &    -       &  0.11(3)   &    -       \\  
 18.56(8)   &    -      & 0.117(5)   &    -       &    -       \\ 
 19.82(9)   & 0.118(15) &    -       &    -       &    -       \\  
 20.40(9)   &    -      &    -       &    -       & 0.1893(27) \\
 21.2(1.1)  &    -      &    -       &  0.12(3)   &    -       \\
 21.94(10)  & 0.108(14) &    -       &    -       &    -       \\
 21.96(9)   &    -      & 0.121(7)   &    -       &    -       \\ 
 24.31(10)  & 0.111(14) &    -       &    -       &    -       \\
 24.57(10)  &    -      & 0.124(7)   &    -       &    -       \\
 26.93(11)  & 0.113(14) &    -       &    -       &    -       \\
 29.10(13)  &    -      & 0.126(7)   &    -       &    -       \\
 29.86(12)  & 0.111(15) &    -       &    -       &    -       \\
 32.58(13)  &    -      & 0.127(8)   &    -       &    -       \\  
 33.13(13)  & 0.104(14) &    -       &    -       &    -       \\
 34.4(3)    &    -      &    -       &    -       & 0.224(5)   \\
 38.60(16)  &    -      & 0.121(10)  &    -       &    -       \\ 
\cline{1 - 5}
\end{tabular}
\end{center}

\newpage

\begin{center}
FIGURES 
\end{center}

\vspace{1 cm}
 
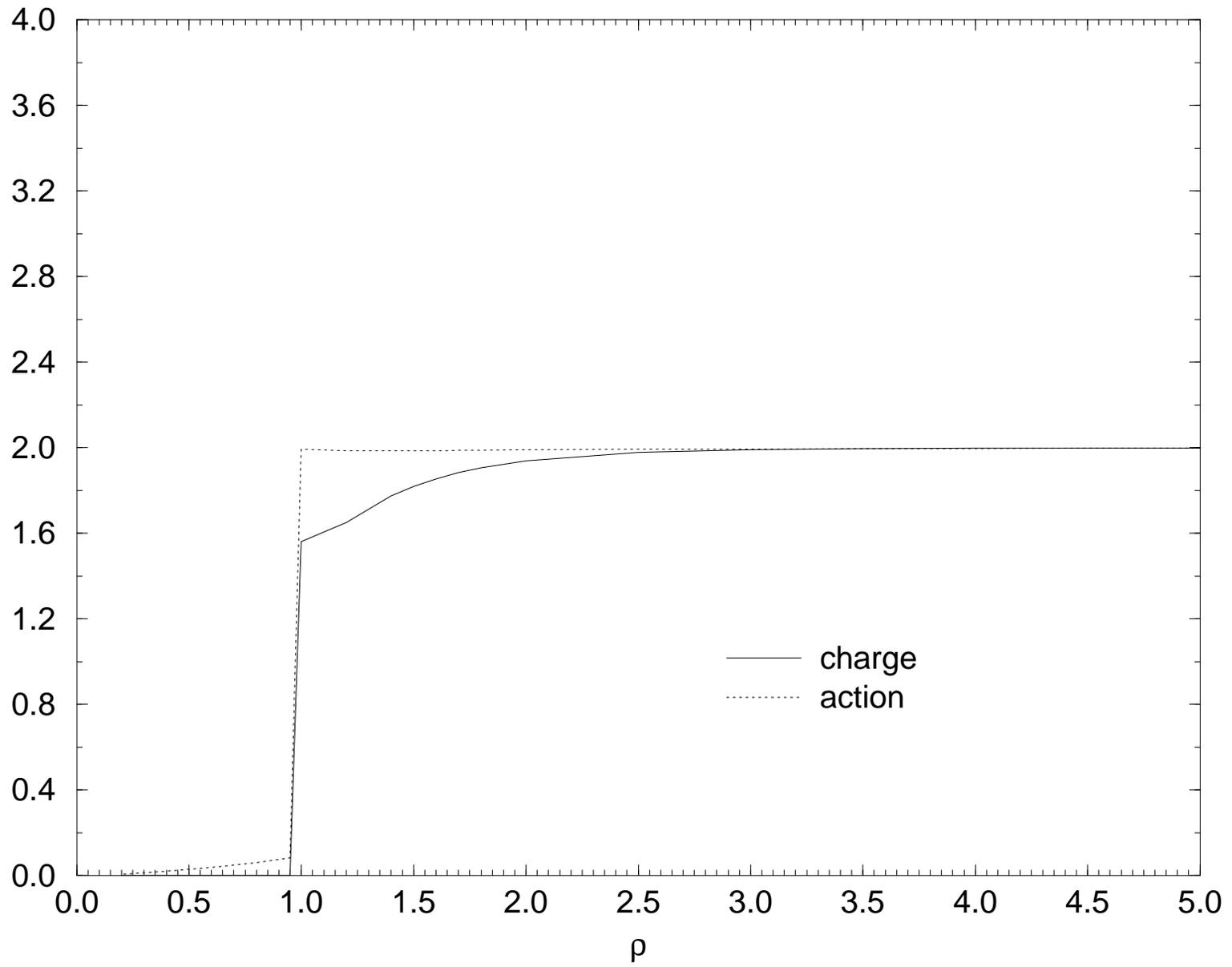
FIG. 1. Cooling of small--size 2--instantons with the FP action.
Action (dotted line) and topological charge (solid line) of cooled 
discretized 2--instantons are represented versus their size in lattice units.
The cooled configurations have been obtained by performing 100 cooling steps 
on the discretized instanton solution. The action is reported in units of 
4$\pi$. The lattice size is 60.
 
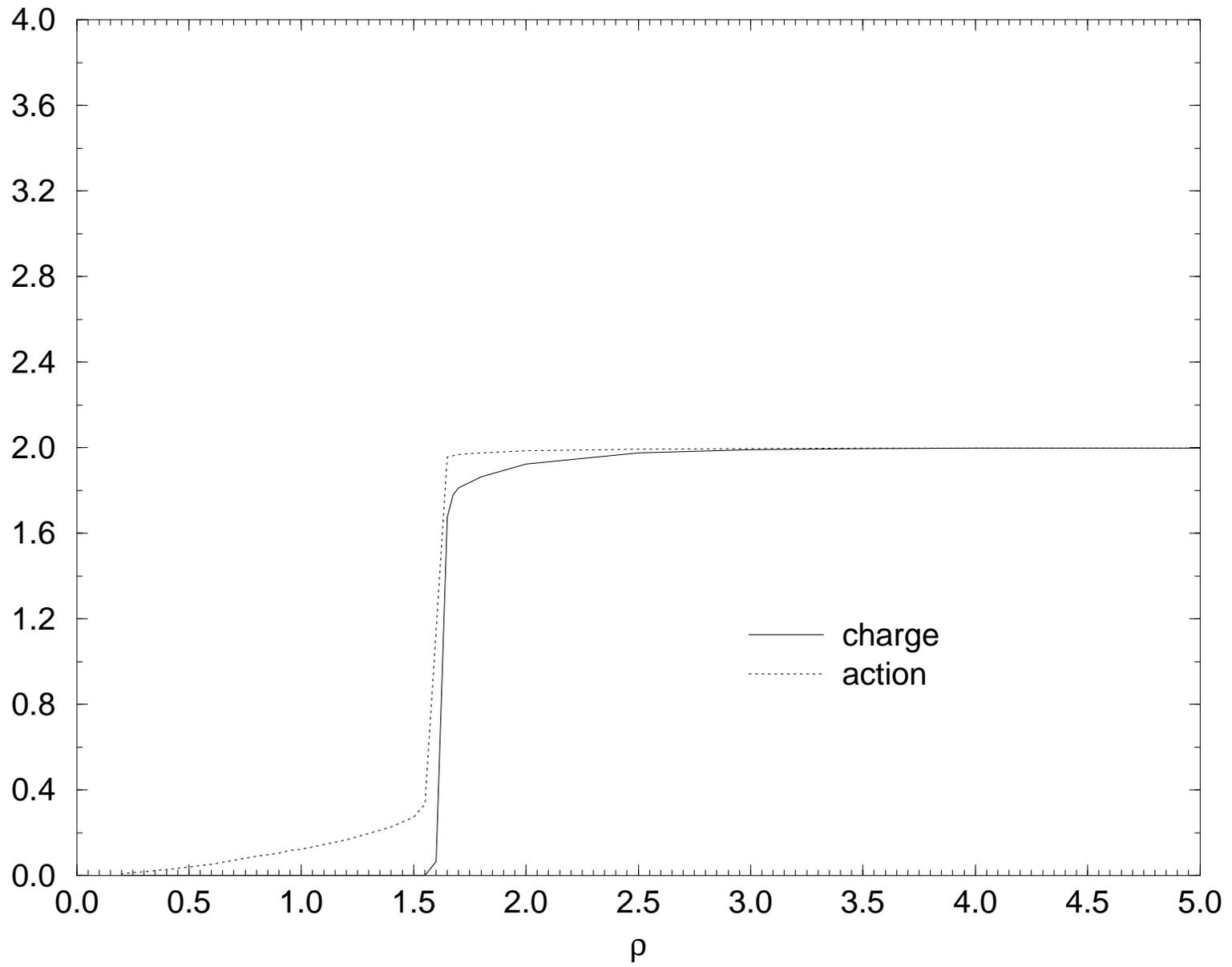
FIG. 2. As in Fig.~1 with the tree level Symanzik improved action. 

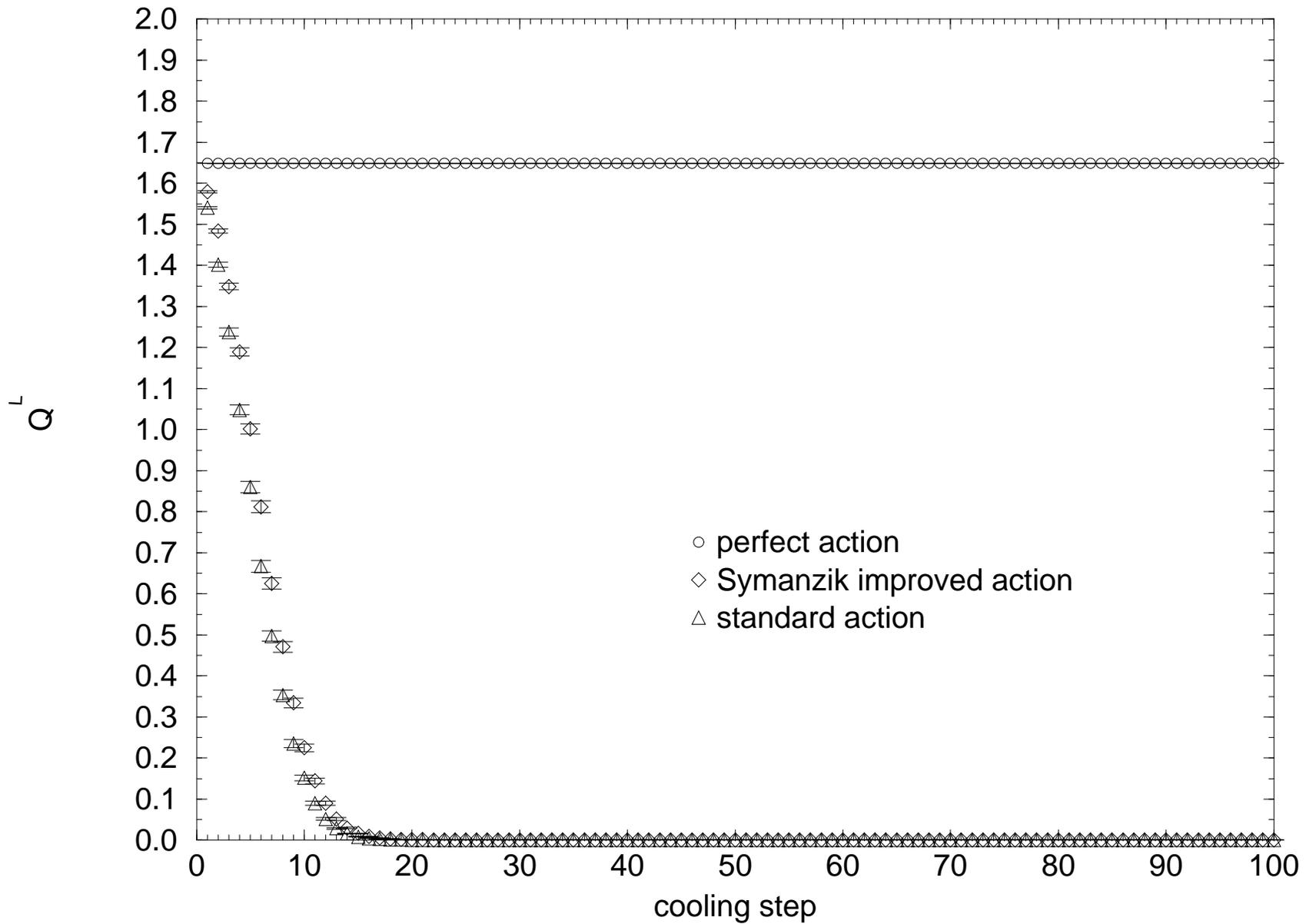
FIG. 3. Stability under cooling of a small--size 2--instanton.
The topological charge of a cooled discretized 2--instanton is represented 
versus the step of cooling when the FP action (circle), the tree level 
Symanzik improved action (diamond) and the standard action (triangle) is 
minimized. The size of the 2--instanton in lattice units is 1.2. The lattice 
size is 60. The procedure has been repeated 100 times with different random 
sequences.

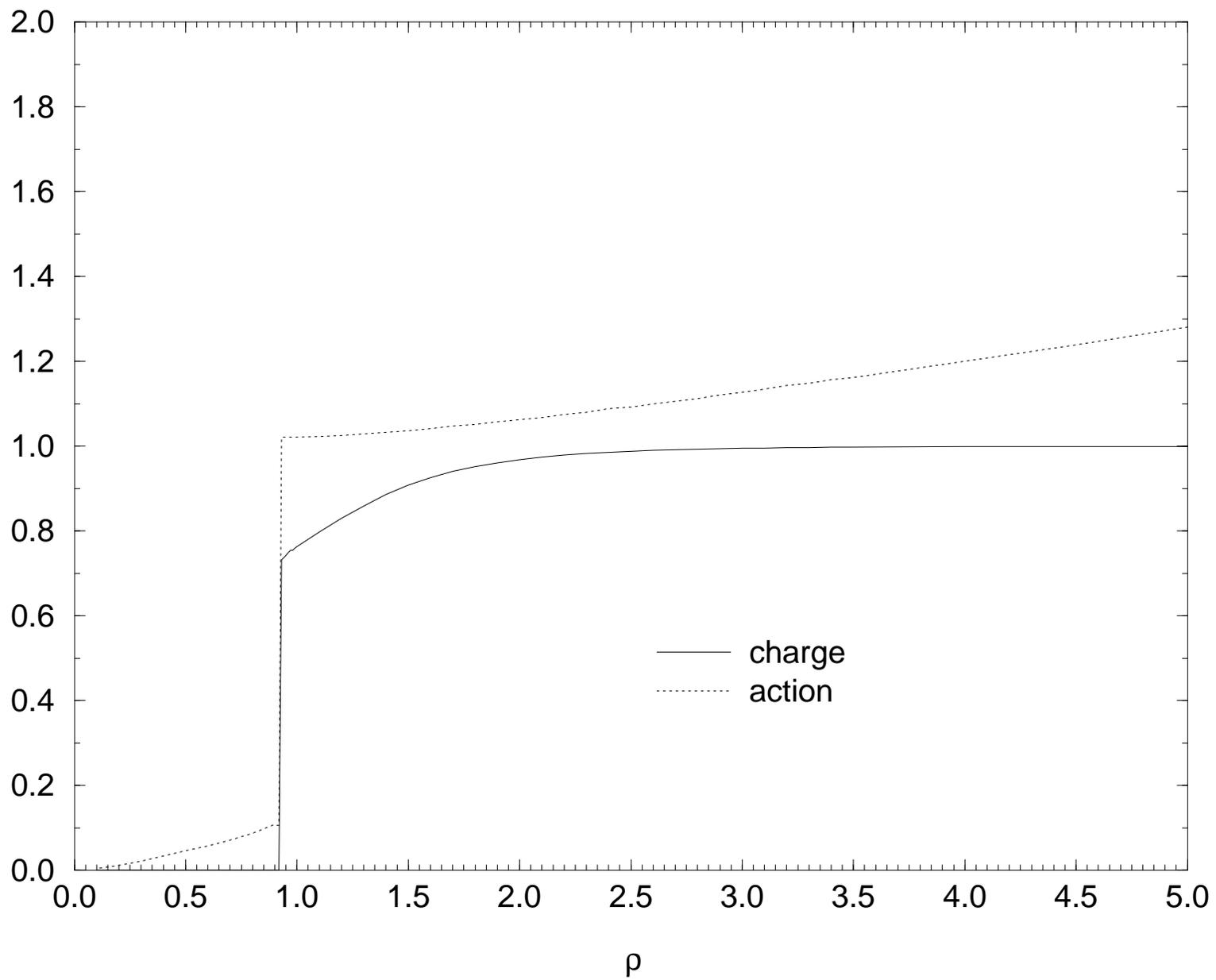
FIG. 4. As in Fig.~1 with 1--instantons (constant boundary conditions). 

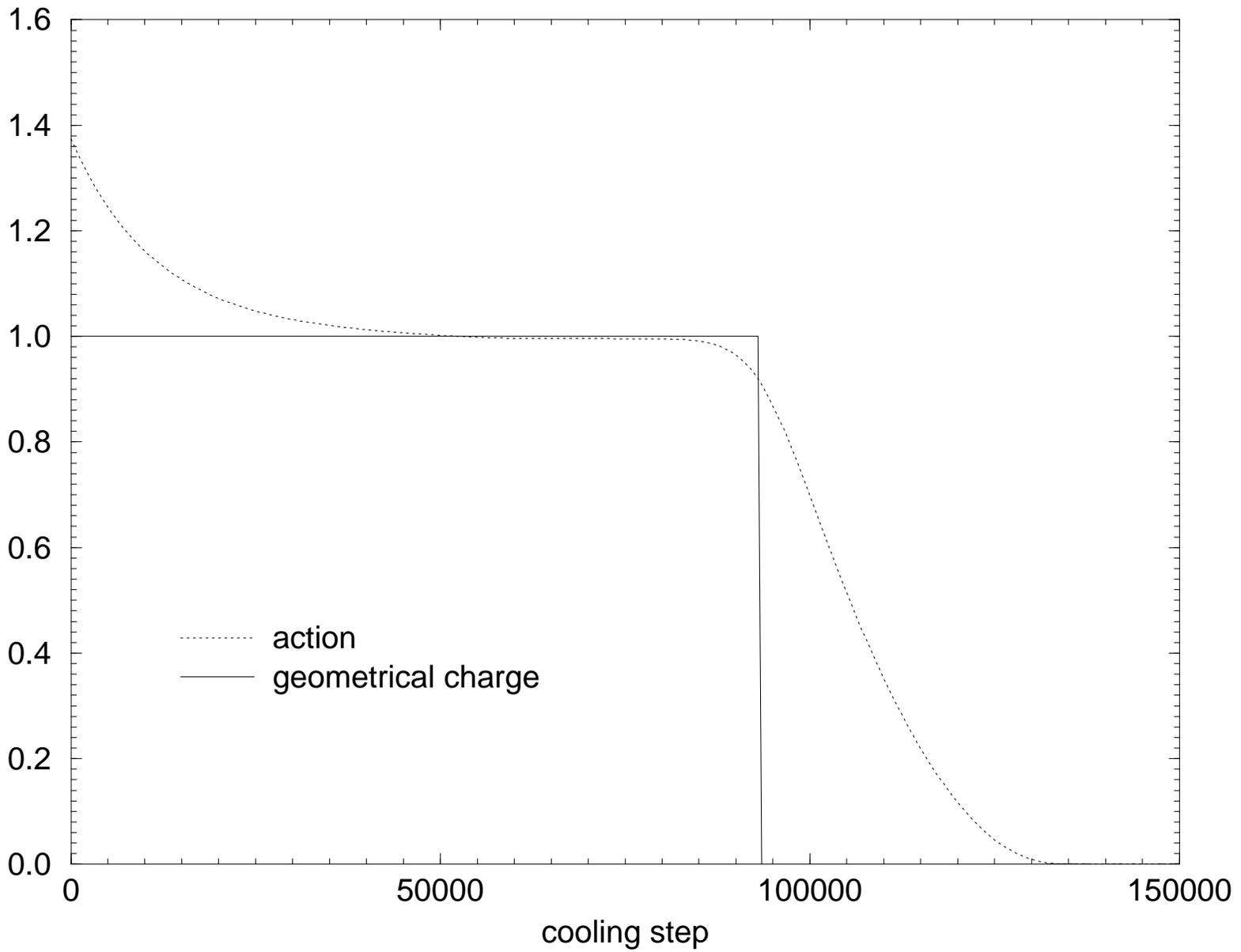
FIG. 5. Transition from $Q_g = 1$ to $Q_g = 0$ during the cooling of
a 1--instanton (constant boundary conditions) with the FP action. 
The behavior of action and topological charge in the geometrical definition 
is represented versus the cooling step. The instanton size in lattice units 
is 15, the lattice size is 60.

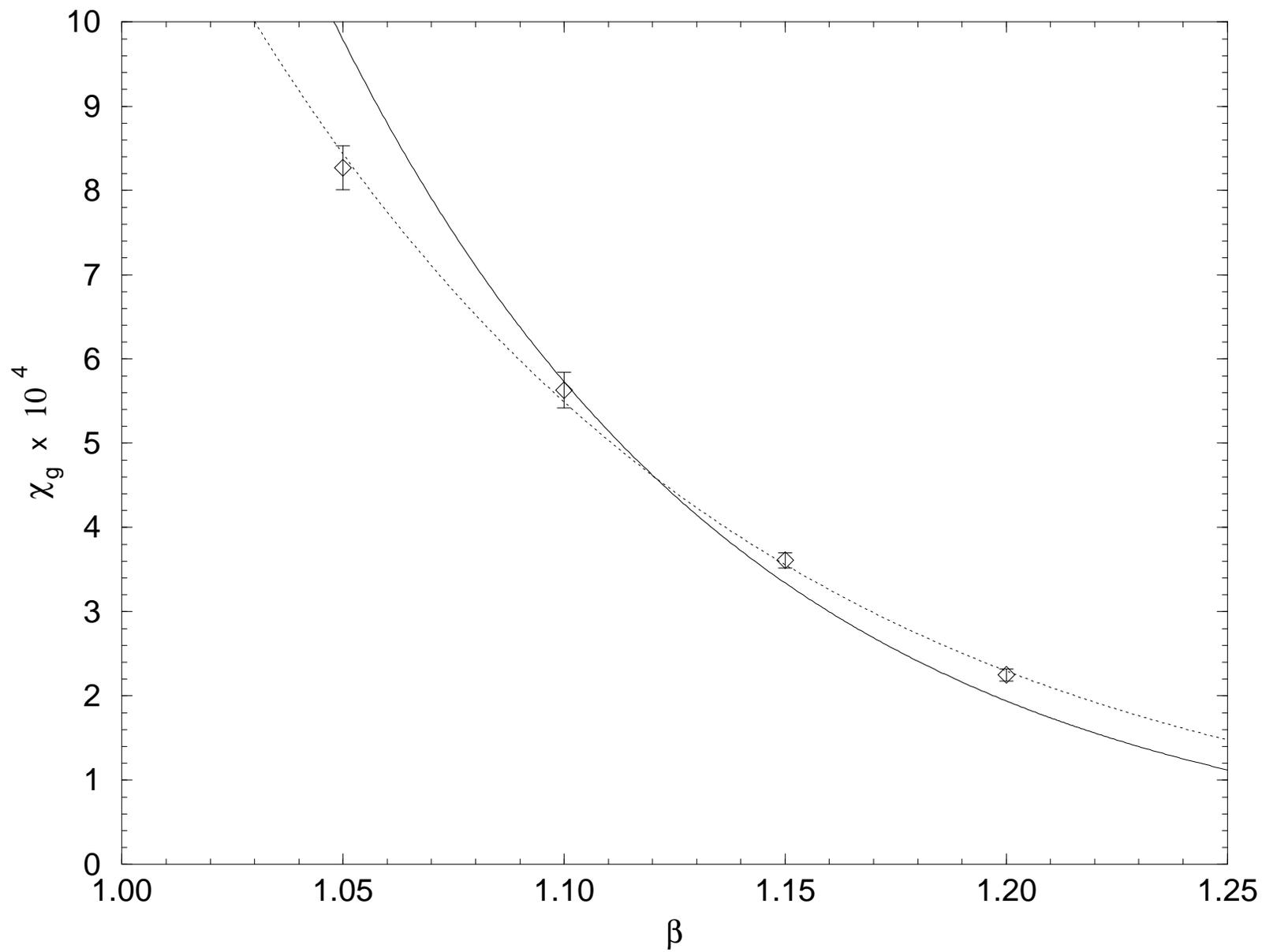
FIG. 6. Scaling behavior of the lattice topological susceptibility in the 
geometrical approach $\chi_g$ with the FP action. Diamonds represent data 
from MC simulations; the dotted line is the best fit, corresponding to an 
exponential decay with $\alpha\simeq0.83$ (see the text); the solid line is 
the scaling behavior predicted by  the renormalization group. The lattice 
size is 60.

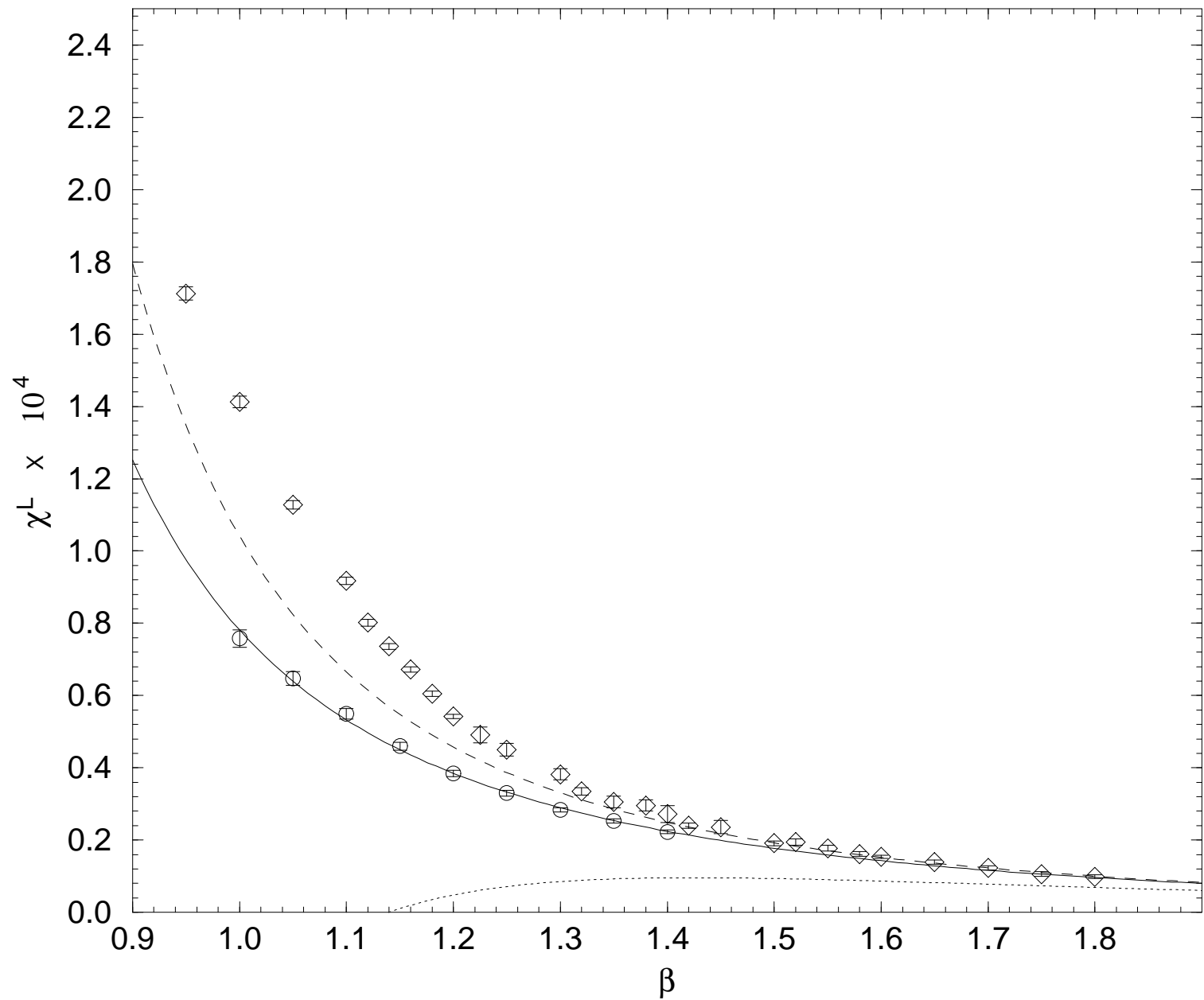
FIG. 7. Comparison between the perturbative and non--perturbative
determination of $P$ -- the mixing of $\chi^L$ with the unity operator -- 
versus $\be$. The diamonds represent the MC data of $\chi^L$, the circles 
the outcomes of the heating technique (the solid line is the relative fit);
the dotted line is the 4--loop perturbative determination of $P$, 
while the dashed line is the best fit of the MC data in the perturbative
region ($\be\geq1.6$).

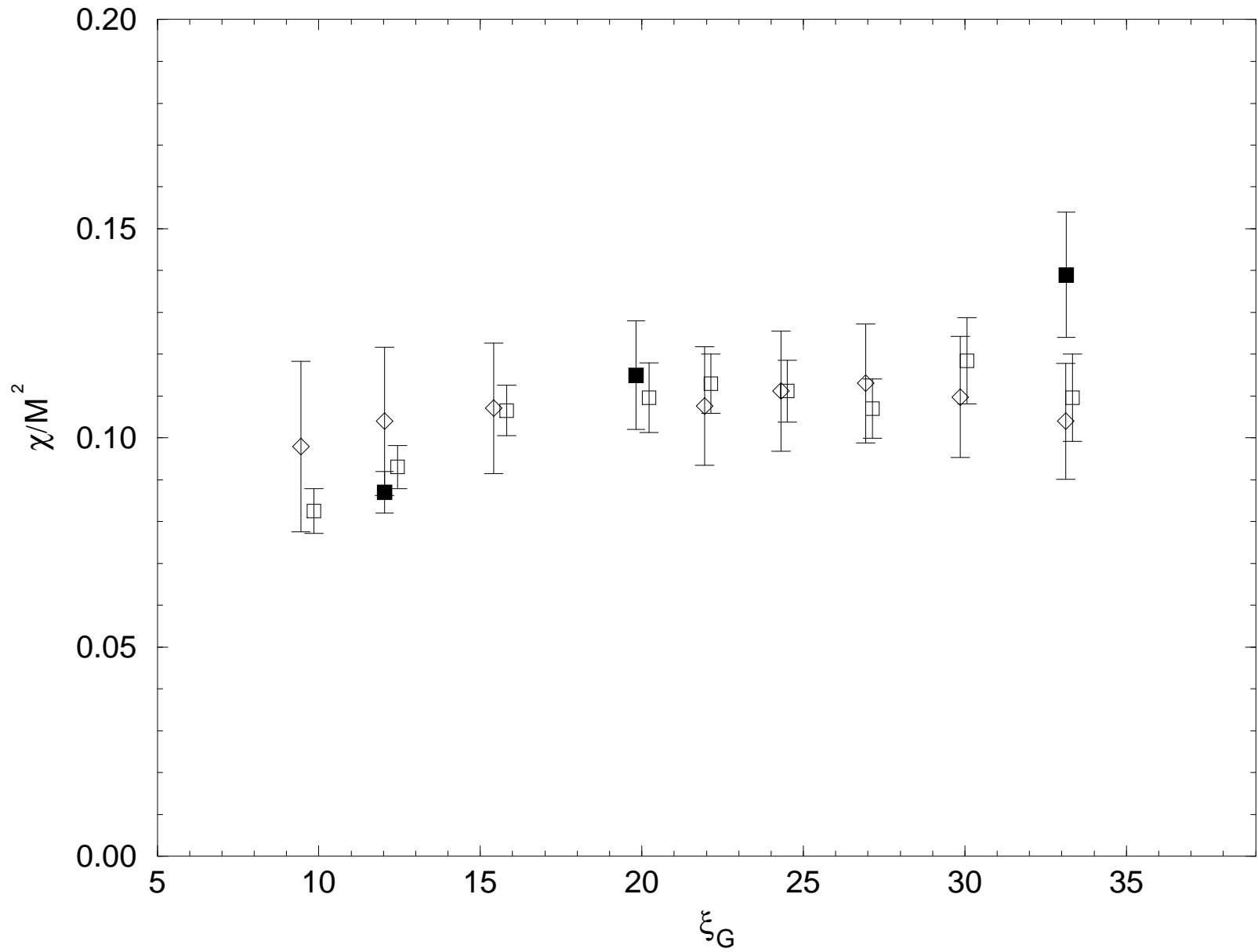
FIG. 8. $\chi/M_G^2$ versus $\xi_G$, the correlation length in lattice units,
with the FP action. \newline
(a) $(\chi^L(\be)-P(\be))Z(\be)^{-2}\xi_G(\be)^2$ on a $60^2$ lattice 
(diamond); \newline 
(b) $\chi/M_G^2$ extracted by cooling on a $60^2$ lattice (square); \newline
(c) the same quantity of (b) computed on lattices of different sizes
keeping $L/\xi_G \sim 6$ (filled square).
For the sake of legibility, data (b) have been slightly shifted in $\xi_G$.


\newpage



\begin{thebibliography}{99}

\bibitem{Sym80} K. Symanzik, {\it in} Recent developments of gauge theories, 
ed. G. 't Hooft et al. (Plenum, New York, 1980); {\it in} Lecture Notes
in Physics, 153, ed. R. Schrader et al. (Springer, Berlin, 1982);  {\it in}
Non--perturbative field theory and QCD, ed R. Jengo et al. (World Scientific,
Singapore, 1983); \\
K. Symanzik, Nucl. Phys. {\bf B226} (1983) 187 and 205; \\
B. Berg, I. Montvay and S. Meyer, Nucl. Phys. {\bf B235}[FS11] (1984) 149; \\
G. Martinelli, G. Parisi and R. Petronzio, Phys. Lett. {\bf B100} (1981) 
485; \\
G. Parisi, Nucl. Phys. {\bf B254} (1985) 58; \\
P. Weisz,  Nucl. Phys. {\bf B212} (1983) 1; \\
P. Weisz and R. Wohlert, Nucl. Phys. {\bf B236} (1984) 397; \\
G. Curci, P. Menotti and G. Paffuti, Phys. Lett. {\bf B130} (1983) 205; \\
B. Berg, A. Billoire, S. Meyer and C. Panagiotakopoulos, Comm. Math. Phys.
{\bf 97} (1985) 31; \\
M. Falcioni, G. Martinelli, M.L. Paciello, G. Parisi and B. Taglienti,
Nucl. Phys. {\bf B225}[FS9] (1983) 313; \\
M. L\"{u}scher and P. Weisz, Comm. Math. Phys. {\bf 97} (1985) 59; \\ 
M. L\"{u}scher and P. Weisz, Nucl. Phys. {\bf B240}[FS12] (1984) 349; \\
B. Sheikholeslami and R. Wohlert,  Nucl. Phys. {\bf B259} (1985) 572.

\bibitem{HN94} P. Hasenfratz and F. Niedermayer, Nucl. Phys.  
{\bf B414} (1994) 785; P. Hasenfratz, Nucl. Phys. {\bf B} (Proc. Suppl.)
{\bf 34} (1994) 3; F. Niedermayer, {\it ibid.} 513.

\bibitem{Wil74} K.G. Wilson and J. Kogut, Phys. Rep. {\bf C12} (1974) 75; \\
K.G. Wilson, Rev. Mod. Phys. {\bf 47} (1975) 773; {\it ibid.} {\bf 55} (1983)
583.

\bibitem{Wil80} K.G. Wilson, {\it in} Recent developments of gauge theories, 
ed. G. 't Hooft et al. (Plenum, New York, 1980).

\bibitem{DHHN95} T. DeGrand, A. Hasenfratz, P. Hasenfratz and F. Niedermayer,
preprint COLO--HEP--361, BUTP--95/14, (1995).

\bibitem{FHNP95} F. Farchioni, P. Hasenfratz, F. Niedermayer and A. Papa,
preprint BUTP--95/16, IFUP--TH 33/95, (1995).

\bibitem{BL81} B. Berg and M. L\"uscher, Nucl. Phys. {\bf B190}[FS3] (1981) 
        412.

\bibitem{Lue82} M. L\"uscher, Nucl. Phys. {\bf B200}[FS4] (1982) 61.

\bibitem{Ber81} B. Berg, Phys. Lett. {\bf B104} (1981) 475.

\bibitem{MPV82} G. Martinelli, R. Petronzio and M.A. Virasoro,
Nucl. Phys. {\bf B205}[FS5] (1982) 355.

\bibitem{BP85} B. Berg and C. Panagiotakopoulos, Nucl. Phys. {\bf B251}[FS13]
(1985) 353.

\bibitem{DFPV92} A. Di Giacomo, F. Farchioni, A. Papa and E. Vicari,
Phys. Rev. {\bf D46} (1992) 4630; Phys. Lett. {\bf B276} (1992) 148.

\bibitem{BBHN94}  M. Blatter, R. Burkhalter, P. Hasenfratz 
and F. Niedermayer, Nucl. Phys. {\bf B} (Proc. Suppl.) {\bf 42} (1995) 799.

\bibitem{AB95}  B. All\'es and M. Beccaria, `The 2--dimensional non--linear
$\sigma$--model on a random lattice', Pisa preprint, IFUP--TH 11/95,
hep--lat/9503025.

\bibitem{FTM} M. Campostrini, A. Di Giacomo and H. Panagopoulos, Phys. 
Lett. {\bf B212} (1988) 206; \\
M. Campostrini, A. Di Giacomo, H. Panagopoulos and E. Vicari,
Nucl. Phys. {\bf B329} (1990) 683; Nucl. Phys. {\bf B} (Proc. Suppl.)
{\bf 17} (1990) 634; \\
M. Campostrini, A. Di Giacomo, Y. G\"und\"u\c{c}, M.P. Lombardo, 
H. Panagopoulos and R. Tripiccione, Phys. Lett. {\bf B252} (1990) 436.

\bibitem{Tep86} M. Teper, Phys. Lett. {\bf B171} (1986) 81 and 86.

\bibitem{WKB} A. Jevicki, Nucl. Phys. {\bf B127} (1977) 125; \\
D. F\"orster, Nucl. Phys. {\bf B130} (1977) 38; \\
B. Berg and M. L\"uscher, Comm. Math. Phys. {\bf 69} (1978) 57; \\
V.A. Fateev, I.V. Frolov and A.S. Schwarz, Nucl. Phys. {\bf B154} (1979) 1.

\bibitem{RR83} J.--L. Richard and A. Rouet, Nucl. Phys. {\bf B211} (1983) 
447.

\bibitem{MS94} C. Michael and P.S. Spencer, Phys. Rev. {\bf D50} (1994) 7570.

\bibitem{FP94} F. Farchioni and A. Papa, Nucl. Phys. {\bf B431} (1994) 686.

\bibitem{BZ-JLeG76} E. Brezin, J. Zinn--Justin and J.C. Le Guillou,
Phys. Rev. {\bf D14} (1976) 2615.

\bibitem{BZ-J76} E. Brezin and J. Zinn--Justin,
Phys. Rev. {\bf B14} (1976) 3110.

\bibitem{BLS76} W.A. Bardeen, B.W. Lee and L.E. Shrock,
Phys. Rev. {\bf D14} (1976) 985.

\bibitem{Z-Z78} A.B. Zamolodchikov and  A.B. Zamolodchikov, 
Nucl. Phys. {\bf B133} (1978) 525; Ann. of  Phys. {\bf 120} (1979) 253.

\bibitem{tH76} G. 't Hooft, Phys. Rev. Lett. {\bf 37} (1976) 8;
Phys. Rev. {\bf D14} (1976) 3432.

\bibitem{DLD78} A. D'Adda, P. Di Vecchia and M. L\"{u}scher, 
Nucl. Phys. {\bf B146} (1978) 63.

\bibitem{BP75}  A.A. Belavin and A.M. Polyakov, JETP Lett., Vol. {\bf 22}, 
No. 10 (1975) 245.

\bibitem{Cre79} R. J. Crewther, Nuovo Cimento, Rev. series 3,
Vol. {\bf 2}, (1979) 8.

\bibitem{CR91} M. Campostrini and P. Rossi,  Phys. Lett. {\bf B272} (1991) 
305.

\bibitem{HMN90} P. Hasenfratz, M. Maggiore and F. Niedermayer, Phys. Lett. 
            {\bf B245} (1990) 522.

\bibitem{DKRV81} P. Di Vecchia, K. Fabricius, G. C. Rossi
and G. Veneziano, Nucl. Phys. {\bf B192} (1981) 392; \\
K. Ishikawa, G. Schierholz, H. Schneider and M. Teper,
Phys. Lett. {\bf B128} (1983) 309.

\bibitem{Tep89} M. Teper, Phys. Lett. {\bf B232} (1989) 227.

\bibitem{DV92} A. Di Giacomo and E. Vicari, Phys. Lett. {\bf B275} (1992) 
429.

\bibitem{CRV92} M. Campostrini,  P. Rossi and E. Vicari, Phys. Rev. 
{\bf D46} (1992) 2647.


\end{thebibliography}
\end{document}